\documentclass[11pt]{article}
\def\baselinestretch{1.5}
\usepackage{graphicx}
\begin{document}
\renewcommand{\baselinestretch}{1.2}
\title{Two time solution to quantum measurement paradoxes}
\author{Michael Grady\\
Department of Physics\\ State University of New York at Fredonia\\
Fredonia NY 14063 USA\\ph:(716)673-4624, fax:(716)673-3347, email: grady@fredonia.edu}
\date{\today}
\maketitle
\thispagestyle{empty}
\renewcommand{\baselinestretch}{1.2}
\begin{abstract}
It is hypothesized that the Langevin time of stochastic quantum quantization is a physical time over which quantum fields
at all values of space and coordinate time fluctuate. The average over paths becomes a time average as opposed to an ensemble average.
It is further hypothesized that the Langevin time also paces the motion of particles through coordinate time 
and is equal to the coordinate time of the 
present hypersurface in the frame of the Hubble expansion. Despite having a preferred frame,
special relativity continues to hold in this
formulation as a dynamical symmetry due to the presumed Lorentz invariance of interactions.
The measurement process becomes an integral part of the theory and is realized as a process of spontaneous symmetry breaking.
The continuously fluctuating {\em history} of fields, characteristic of having two times, and the switch from 
ensemble averages to time averages allows for logical and straightforward explanations of many quantum measurement paradoxes. 
The fluctuating history also evades hidden-variable prohibitions allowing an essentially classical system 
to underlie quantum mechanics.
These changes to the stochastic quantization paradigm makes this stochastic classical system
differ somewhat from standard quantum mechanics, so, in principle, distinguishable from it.
 
\end{abstract}


\newpage
\section{Introduction}
Simulations of quantum field theory using Monte Carlo, molecular dynamics or Langevin methods all have a second time over 
which configurations are generated. This ``Langevin time" or ``Monte Carlo time" is usually thought of as simply a part
of the technique, with no physical significance.  It is used to generate field configurations that are then used in ensemble 
averages.  However, to the extent these techniques generate correct answers, for which much evidence exists, one is led 
to the possibility that the physical world is described more directly by these equations than the original quantum field 
theory from which they were developed.  As Langevin time advances, fields at all
values of space and coordinate time fluctuate. If one additionally 
hypothesizes {\em some} connection between the  Langevin time and the compelled 
motion of particles through
coordinate time, where the Langevin time in some sense paces the motion through coordinate time,
then one obtains a system that has a lot in common with quantum systems but differs in some useful aspects. 
It is possible these differences, though in principle observable, are so small as not to have been observed.  In this case
the real world may more closely resemble the two-time Langevin system than it does standard quantum mechanics, so
it is worth considering it as an alternative.  In fact, it seems to have many advantages over standard quantum mechanics
in that the measurement process is easily incorporated into the theory itself, and most quantum ``paradoxes" involving 
superpositions and entanglement appear to
have simple non-mysterious explanations.

These are classical but stochastic systems that fluctuate in the 
Langevin time from random inputs. Expectation values become time averages of these rapidly varying classical configurations, which
have definite values at any actual instant of Langevin time. 
The idea is to replace
the ensemble averages of quantum mechanics with time averages over fluctuating classical systems. These are not exactly the same,
because the time average is only over the time of an observation, rather than an infinite time, so it is a truncation from
the full ensemble average, which could have observable differences, for instance for rapidly sequenced observations.
This scenario also allows one to understand how
spontaneous symmetry breaking can practically take place in a finite system as an historical event and how measurements happen 
in quantum mechanics through this process of spontaneous symmetry breaking. The fact that the past continues to exist and 
is still fluctuating allows one to explain systems such as the double slit, EPR experiment, entanglement, and Schr\"{o}dinger's cat 
with simple logical explanations that are no longer mysterious. The measurement process is brought completely within 
quantum mechanics with no additional hypotheses or interpretations required. There is also no conflict with relativity so
long as the Lagrangian is Lorentz invariant, although relativity from this point of view is more a consequence of dynamics 
than kinematics.

Although usually the above methods are performed in Euclidean space, the Langevin approach to stochastic quantization\cite{parisi-wu} 
can at least in principle be 
taken directly in Minkowski space\cite{sqmin} which is the version one would need if the simulation were to be directly equivalenced 
to reality.  This scenario still leaves open the source of fluctuations, which are simply postulated to be random variables
located at each spacetime point that interact with the fields.  It also leaves out the sticky question of ``the present" and what
determines it, and why we seem to be compelled to move through coordinate time. This problem it shares, of course, with 
ordinary quantum field
theory. Physicists are divided on whether the lack of a physics explanation for the existence of ``the present" is a flaw
in present-day theories\cite{muller}.

A more radical two-time approach that solves these problems while also explaining the source of fluctuations 
is the Phase Boundary Universe proposal \cite{pbu}.  This
is a classical model that has four spatial dimensions and one Newtonian universal time.  The space is filled with a supercooled liquid at
a certain temperature which undergoes ordinary thermal fluctuations.  A nucleation event starts a crystal growing which is the big bang.
Our 3-d universe is located at the surface of the crystal - the growing phase boundary, which we view as ``the present."  The fourth
spatial dimension that the crystal is growing into is interpreted by us as a time coordinate because we are compelled to move
through it due to the growth of the crystal, which creates the Hubble expansion.  Here the source of fluctuations are simply the 
ordinary thermal fluctuations of the 4-d substance (crystal + liquid) which take place in universal (Langevin) time.  
Elementary fermions might arise as crystal dislocations and 
bosons as phonon-like excitations confined to the surface, and perhaps even gravity as a surface tension effect. Special relativity
arises from the dynamics. This universe is not based on a vacuum solution for the background space but rather
on a non-equilibrium dynamic phase front with reduced symmetry, the symmetry of our universe. There is a preferred frame, 
the co-moving frame of the expansion, but it is not detectable
until inverse momenta are small enough to approach the crystal spacing. A ``world-crystal" model for the universe with 
some similarities to this has also been
proposed by Kleinert and Zaanen\cite{kleinert}.  Unfortunately a satisfactory model resembling our universe is 
still a long way off in the phase boundary approach. It may be required to replace the crystal  with something more 
exotic such as a liquid crystal or one of the phases 
found in He-3\cite{volovik}.  Nevertheless, solving the mystery of the present (also addressed in \cite{muller}) and 
how it differs from past and future,
explaining quantum fluctuations as thermal fluctuations, giving a reason for the big bang and Hubble expansion, and very
possibly reconciling quantum mechanics with gravity are compelling reasons to work within this framework. Due to the two times,
the quantum measurement ideas presented here also hold in the phase boundary universe model, in fact that is their
origin.  However, because the quantum measurement results are simply a consequence of the two times, this paper is more
generally cast for any theory that achieves stochastic quantization with a second time. The one which presents the least radical 
departure from standard quantum field theory is the Langevin scenario depicted above, so that is the system that
will be mostly considered here.

\section{Quantum measurement process as spontaneous symmetry breaking}
The quantum measurement process has always been a chink in the armor of quantum mechanics.  The Von Neumann 
wave function collapse that 
occurs in the standard Copenhagen interpretation is a process beyond the operation of the Schr\"{o}dinger equation, and 
although mechanisms that involve added interactions have been suggested\cite{ghirardi}, there is no 
widely agreed upon equation that describes how it takes place. In addition, there is no agreed-upon definition of what 
constitutes a measuring device or a clear operational distinction between a quantum object which can exist in a superposition 
and a measuring device which 
presumably cannot.  The measurement process has a close similarity to the process of spontaneous symmetry breaking(SSB). When
a quantum system undergoes SSB one always ends up in a single definite ground state with a specific
value of the order parameter, never in a superposition of several ground states. Although that would seem to be possible from normal
quantum evolution it is ruled out by superselection rules or similar arguments forbidding it. This looks very much like 
wavefunction collapse.  For instance at the electroweak symmetry breaking in the early universe when the Higgs particle gains a mass,
one can think of the universe in a sense measuring itself for the Higgs field to fall into a specific vacuum. Superselection rules
can come and go at a symmetry breaking. For instance, normally there is a superselection rule prohibiting the 
superposition of states of different electric charge, however if the electromagnetic symmetry breaks spontaneously, 
as in the abelian Higgs model, the vacuum itself has this property. The usual argument for existence of a superselection
operator is if there is no interaction connecting the states, there is no way to ever produce the superposition. 
This is usually a result of symmetry.  In a classical statistical mechanical system operating
in time (e.g. microcanonical) undergoes SSB, due to a lowering temperature for instance, then it is clear how it gets stuck in
a particular sub-ensemble simply by accident by virtue of the particular state it was in when the temperature fell below the 
transition temperature.
Providing the system is infinite, it will never make the transition to another sector - they are no longer ergodically connected.
The situation is less clear in the canonical ensemble where all configurations are still counted, however adding a small
external field allows one to choose a ground state in this formalism.  For a finite system, there is technically 
no SSB. It still occurs for all practical purposes in finite
systems if the tunneling times between different ground states becomes long compared to the observation 
time (such as the age of the universe).  Thus there are significant differences between the canonical ensemble where all
configurations are used and the time-series of states of a single system, averaged over a large but finite time\cite{microcanonical}.
This can be either a microcanonical or Langevin evolution.

Given the close similarity of wave-function collapse in measurement and choice of ground state in SSB,
it is tempting to try to model all measurements as a process of SSB\cite{qmasssb}. Measuring devices may be pictured as machines that can
exist in two phases, one spontaneously broken and one not, with a knob that can be turned that adjusts the parameter that
causes the symmetry to break. The device is then coupled to the property of a quantum system one wishes to measure, and then the 
symmetry breaking knob is
activated, carrying both the measuring device and the quantum system, now strongly coupled to it, into a single broken state.
However, one is again faced with problems in the canonical formulation if the measuring device is not infinite, because
then the ensembles are not fully bifurcated, superselection operators not exact and the normal problem of the measuring 
device itself ending up in a superposition ensues.  Cosmological symmetry breaking in a finite universe suffers 
from the same problem. Both are solved if one switches to a stochastic system that varies over (Langevin) time, taking time 
averages instead of ensemble averages. In this case symmetry breaking does take place in a finite system for all practical
purposes, just as it does in a finite piece of a crystal or magnet, because one is not going to 
observe it for an infinite time.

There are some statements in the literature that decoherence removes the need for wave function collapse\cite{decoherence}, 
but many others feel
that this is only a partial solution\cite{adler}.  It does not seem to help with the cosmological phase transition, for instance.  
Here I
take the point of view that wave function collapse is a necessary feature of quantum mechanics which requires an explanation.

\section{Double slits and interferometers}
Let us consider the electron double slit experiment from this perspective.  In the usual picture, the fact that an 
interference pattern is still created if electrons are sent at the slits one at a time, leads to the conclusion that each
electron passes through both slits.  The complementarity principle allows us to accept this, because, as has been shown many times,
if one observes which slit the electron goes through the pattern is then destroyed.  Nevertheless it is unsettling to
imagine an electron in vacuum actually splitting in two and going through both slits, considering it has proven to be impossible 
to split 
an electron in half by force even given the huge energies available in particle accelerators.
If it is not ``exactly'' splitting in two then how ``exactly'' is
it going through both slits?  In a two-time theory,  past values of the fields continuously fluctuate in Langevin time.
For a single electron propagating the field has non-zero occupation number only for 
present and past, since there is no electron yet in this vicinity in the future.
This means that the electron world-line drifts around as it is also extended --- history is constantly being rewritten as 
time progresses (the coordinate time
of the particle's and observer's present and Langevin time being linked).  At one instant the electron 
``has" passed through one slit, perhaps just having arrived near the screen and 
wandering around (not yet having been absorbed). An instant later the electron world-line trajectory wanders to the second slit
which is able to reconnect with the world line on the other side (perhaps similar to magnetic reconnection). We are still 
considering that the electron has a phase factor associated with it, i.e. a wavelike nature.  As the electron is hitting the screen,
still many instants of Langevin time occur before it is absorbed. Thus it wanders around from spot to spot on the screen,
undergoing phase angle changes as well, due to path length differences, sometimes 
going through one slit and sometimes the other. Say
it is a photographic screen which will initiate an irreversible chemical reaction if a certain activation energy is deposited. This
energy must be built up over some time. Each time the electron returns to that spot if the phase angle is close to the same as before
the effect will add,
if opposite it will subtract from the previous interaction and possibly cancel it.  These effects are averaged over a certain
time scale characteristic of the absorber, possibly short on the human scale, but long on the Langevin scale.
Once actually absorbed the electron is captured in a molecule and that end of the trajectory
is then tied down.  This is because so many degrees of freedom have taken place in the irreversible reaction that the transverse 
translation
symmetry the electron had is now spontaneously broken. Its position at the final time is nearly definite, the remaining fluctuations
being much smaller due to the large numbers of coupled degrees of freedom, and the dissipation of energy. 
This is also what happens if one monitors the slits. Coupling large numbers of degrees of freedom to the electron at the position of
the slit breaks the translation symmetry {\em there} preventing fluctuations from ``rewriting history" to the extent of
switching slits.  Now the
fluctuations at the screen don't include the drift to the other slit, so of course the pattern seen is different.

Every field at each spacetime point changes with each instant of Langevin time. Whether the process is continuous or not is largely 
irrelevant. Symmetries and conservation laws as well as the presence of broken symmetries provide some order in this
chaos (one could argue the {\em only} order). Whatever the exact mechanism of evolution in Langevin time, it must be possible
for large changes to happen quickly because in some cases
the possible paths the electron could be on are widely separated,
such as in the Stern-Gerlach experiment or Mach-Zehnder interferometer. 
So long as symmetry and conservation laws allow it, we are picturing that the system
has a way of moving quickly between allowed configurations that  may be widely separated, {\em provided the number of degrees of freedom that
are changing is still small}, such as the fields along a single electron's world-line. Fields along the world line at 
adjacent time coordinates are highly correlated, so do not count as fully separate degrees of freedom (they must fluctuate together).
This lepton number conservation keeps the number of degrees of freedom along an electron's world-line in the 
relatively small category. Collective degrees of freedom involving many elementary fields, so long as the collective 
degrees of freedom
themselves are 
few in number can 
in principle evolve
quickly in a stochastic system. 

\section{Schr\"{o}dingers cat and quantum entanglement}
What is being proposed here is essentially Feynman's sum over histories, but with the histories being presented one by one ordered
by the Langevin time. This likely gives different results than a complete sum over histories,
because not all histories will be visited in a finite time. The time
for an observation to take place is the relevant time over which the histories are effectively summed. If a large number of 
moderately-coupled degrees of freedom exist in a broken-symmetry state, separated from other such states by an energy or entropy barrier,
then there will likely be insufficient time to bridge these barriers, and the system will be essentially frozen in this state.
This allows for symmetry breaking to take place in finite systems and effectively prevents measuring instruments that rely
on broken symmetries from existing in superpositions.  This is the mechanism's explanation of the Schr\"{o}dinger's Cat 
paradox. The radioactive decay effects too many degrees of freedom to be able to fluctuate back to the undecayed state 
in a reasonable Langevin time. Needless to say, adding the cat's degrees of freedom makes this even less likely.  
Tunneling times generally increase exponentially with system size. So
although there will be a range of mesoscopic systems which will fluctuate slowly and perhaps observably, as the number of degrees of
freedom is raised one quickly moves from a regime of extremely rapid tunneling to one with extremely rare, essentially 
nonexistent tunneling as the number of moderately-coupled degrees of freedom is increased.  
The idea of moderate coupling is an important one in that it involves counting the effective degrees of freedom. 
Two degrees of freedom that are so strongly coupled that they must fluctuate together are essentially only one degree of freedom.  
One can in principle have rather large systems that still have relatively few degrees of freedom, such as for instance a 
Josephen junction, which can still display entanglement.

Fluctuating histories also give an explanation of quantum entanglement in the Einstein-Podolsky-Rosen\cite{epr} gedankenexperiment.
Here a spin-0 particle decays into two spin 1/2 particles moving in opposite directions, the spins of which are perfectly 
oppositely correlated, the two particle wave-function being $(1/\sqrt{2})|\uparrow \downarrow >+(1/\sqrt{2})|\downarrow \uparrow >$.
If the one particle's spin  is measured along a particular axis, giving a result of $\pm \frac{1}{2} \hbar$, then one knows 
immediately the spin component of the other particle along the same axis. Bell's
inequality\cite{bell} violation, confirmed by experiment\cite{aspect}, shows that the result is not 
predetermined as it would in a classical hidden-variable theory, but rather follows the predictions of quantum mechanics.  
The oddity of this example is that the 
fact that one particle is measured seems to instantaneously affect the other particle, which may be far away.  
It does appear to be impossible to use this to send a superluminal message\cite{noFTL}, so it is not really a paradox, but 
still, as described by Einstein, a ``spooky action at a distance.''
In the two-time interpretation, the spins of the two particles along their entire historical world lines drift in concert with each other
in order to preserve angular momentum conservation.  This ``fluctuating 
history" avoids the difficulties of the classical hidden variable model, which can be traced to the 
fixed definite (but simply unknown) history assumed there. When one particle's spin
is measured its spin stops fluctuating, because it is now highly correlated with the large numbers of d.f. of the measuring device.
This fixes the particle's history and also stops the other particle from fluctuating, instantaneously in the Langevin time, because it 
too is correlated with the distant measuring device. Because both spins at all times
are updated at every moment of Langevin time, their 
fluctuations only limited by conservation laws and SSB, this behavior seems rather straightforward. Some time ago Ne'eman 
envisioned the two EPR particles being connected through a rigid connection that traced the two world lines through their source, 
something akin to a differential gear that allowed them to fluctuate only in a correlated manner\cite{neeman}. Our fluctuating
history that respects conservation laws essentially implements this vision.

\section{Interaction free measurements}
Interaction free measurements are another area where quantum mechanics seems to have a non-classical and surprising result. 
At first 
glance one might think the stochastic approach presented here might run into trouble in such a system, but in fact 
the two-time model again has no difficulty explaining it.  A famous case is the Elitzur-Vaidman bomb-testing scenario\cite{ev}.
It is supposed that one has some bombs that have hair-trigger detonators each attached to a mirror.  Some of the bombs have their
hair triggers jammed (duds).  By placing these mirror triggers in one leg of a Mach-Zehnder interferometer one can 
some of the time detect active bombs without setting them off.  The interferometer is arranged so that the recombined beams 
always exit
the final beamsplitter in one of two possible directions (say A).  However, this requires a rigid mirror (dud) so that each photon 
can go through both legs
without being detected, interfering constructively in direction A and destructively in direction B. 
However, if a live bomb is in place then that will measure which leg the photon went through by either
exploding  if the photon hit it or not, in which case the photon took the opposite leg.  But now there is no 
interference because there is no
second beam and the exiting photon can take either path at the final beamsplitter. 
If it takes path B then one knows a bomb is present even without having  interacted with it.  
In the two-time model, as coordinate time progresses the photon worldline rapidly cycles between the two 
interferometer legs as it is extended.  If it hits a fixed mirror the path is extended beyond the mirror.  
For the dud bomb case eventually the paths come together and interfere by the rapid fluctuation between paths mechanism, 
averaging over time, with destructive interference along one route (B) and constructive along the other (A).  Finally it is
detected along the A-leg of the beamsplitter output where the interference is constructive. If a bomb is present,
then there are two possibilities.  If the beam happens to be on the bomb side when it reaches the mirror, it may well transfer
enough momentum to it to trigger the bomb. This will correlate the photon with the many degrees of 
freedom of the moving mirror and exploding bomb preventing the history to fluctuate back, and the beam will never be 
able to switch to the other side again.  However this goes both ways. If the beam was on the non-bomb 
side when it hits that mirror, it can no longer fluctuate back to the bomb side where
it would have passed the bomb mirror.  This would require moving the bomb mirror forward to accommodate the required history
of the photon, but this would be too many degrees of freedom to move simply through fluctuation.  So the non-measurement is a
just as definite fix to the history as a measurement would have been, given only two possibilities. The photon on the non-bomb
side can never fluctuate back to the bomb side after this because there is no longer a history on that side consistent with 
momentum conservation etc., which involves changing only a few degrees of freedom.

\section{Modifying stochastic quantization to implement the two-time relationship}
Stochastic quantization of quantum systems and field theory has been explored with a number of somewhat different
formalisms \cite{stochastic,otherstochastic,kleinert2,mueller} mostly based on the Parisi-Wu ideas\cite{parisi-wu}. 
An earlier attempt by 
Nelson\cite{nelson} differs somewhat in using only a single
time coordinate over which particles move and fields fluctuate.  In a sense the approach here bridges the difference 
between the Nelson approach and later approaches by
using a second (Langevin) time, but still relating the growth of coordinate time to it (developed further below).  
Clearly our approach differs from all
of these approaches, in that the modern approaches to stochastic quantization, which are believed to be fully 
equivalent to quantum mechanics, use a
Langevin time wholly separate from the coordinate time.  However there is one ingredient missing from these modern 
approaches to stochastic quantization - one way in which these classical stochastic systems are not fully equivalent to
the quantum systems they model, and that is a lack of measurement process or theory.  The quantum measurement theory must 
still be applied to results obtained with stochastic simulations.  As succinctly stated by Haba and Kleinert, 
``It remains to solve the open problem of finding a classical origin of the second important ingredient of quantum theory: the
theory of quantum measurement associated with a stochastically evolving wavefunction\cite{kleinert2}."  As outlined above, 
relating the two times allows the measurement theory to become a part of the stochastic system evolution.  Beyond being a possibly 
more accurate description of nature than quantum mechanics itself (the two theories being only approximately the same), 
this also opens the possibility of obtaining the speed of 
algorithms promised by a quantum computer to be obtained by a classical computer with stochastic inputs.  Already quantum
systems have been successfully modeled by stochastic analog computers utilizing noise generators\cite{asq}. By simulating measurement 
processes as well, it seems possible that such a system could fully simulate, i.e. replace, a quantum
computer.  The problems of decoherence and scaling would likely be much less of a factor with such an approach.

Here the proposed method will be demonstrated with a scalar field theory. A scalar field $\phi (x,t)$ with action $S(\phi)$ is 
described by a Langevin equation
\begin{equation}
\frac{\partial \phi(x,t,t')}{\partial t'} = i\frac{\partial S(\phi (x,t))}{\partial \phi (x,t)}\bigg|_{\phi (x,t)=\phi(x,t,t')}
+\eta (x,t,t')-\epsilon \phi (x,t,t'),
\end{equation}
where $\eta (x,t,t')$ is a random Markov noise characterized by
\begin{equation}
<\eta (x,t,t')>=0, <\eta (x_1 ,t_1 ,t'_1 ) \eta(x_2 ,t_2 ,t'_2 )> = 2 \hbar \delta^{n} (x_2  -x_1 )\delta (t_2 -t_1 )\delta (t'_2 -t'_1 ).
\end{equation}
Here $t$ is coordinate time, $x$ is the spatial coordinate ($n=$1-3 dimensions), and $t'$ is the separate Langevin time. 
This is the Minkowski space version with the $\epsilon$ term added for convergence.  In the standard stochastic quantization method
correlation functions are given by averages over different noise sets $\eta (x,t,t')$ and with the Langevin time taken to infinity.
Alternatively by the ergodic theorem one can just average correlation functions over the Langevin time
 once a sufficiently long time has passed for the 
probability distribution to achieve its limiting form. The modification being suggested here is to equate 
the Langevin time to the coordinate time of the 
present hypersurface in a frame to be discussed below. The simulation is run for a sufficient time {\em before} the 
experiment takes place to have equilibrated. Then a measurement when $t'=t_1$ (i.e. when $t_1$ is the present time 
or ``at time $t_1$") 
prepares the initial state. The simulation continues in $t'$ and a second measurement takes place when $t' = t_2$ 
shortly after which the simulation is terminated. Note that the entire
history of the events that took place subsequent to the final measurement is still fluctuating during the second
measurement because all fields at all times fluctuate over the Langevin time. 
This allows the constructive and destructive interference 
to appear through the time average of fluctuating fields over
the characteristic time of the second measuring device, as envisioned above.  
The limit of taking the Langevin time to infinity is not taken, because the measurement will be read shortly after taking it 
and one purposely does not want to wait a very long time 
over which the measuring device could eventually fluctuate to a different state, since it is a finite size.  
This avoids the measuring device from itself ``entering a superposition.''  Instead the Langevin time in principle 
starts at negative infinity - in practice a sufficiently early time so as to forget the very initial state.  A single run 
of the simulation is all the universe itself is doing, of course. However
in order to build up a probability distribution of possible results, the average over different noise sets would also be performed,
but this is at the level of probability and not probability amplitude because it is a post-measurement average.

The modeling of specific measurements as part of the quantum calculation seems unusual, but that is exactly what is
needed if the measurement process is to be brought into quantum mechanics.     
If specific measuring devices are not being modeled, as is more normally the case in quantum mechanical calculations, 
one can simulate their effect by averaging
the correlation functions measured with the simulation over the expected characteristic time of a measurement, $\Delta $. 
For each Langevin simulation run one would compute the expectation value of a correlation function using 
\begin{equation}
\int_{t_2 -\Delta /2}^{t_2 +\Delta /2}\phi (0,t_1 ,t')\phi(x,t',t')dt' \label{this}
\end{equation}
This allows for the ``in-time'' superposition or cancellation referred to above to take place. Then these results are further averaged
over different noise sets $\eta (x,t,t')$ only {after \em} a probability quantity is computed (such as by squaring) to then obtain 
the full probability distribution.

The theory presented here, therefore, differs in two important ways from standard quantum field theory under stochastic quantization.
The first is relating the Langevin time to the coordinate time, with the Langevin time being the coordinate time of the present
hypersurface. The second difference is that the sum over paths that creates interference is limited to 
those that are visited in the relatively short
time $\Delta$ over which a measurement takes place.  The results of the measurements from different runs of the simulation may then 
be gathered to form a probability distribution.  Whether this theory will be easy to distinguish from quantum mechanics
depends on the ratio of times between the characteristic measurement time $\Delta$ and the 
characteristic time of quantum fluctuations
themselves, $\delta$ over which fields change their values significantly.  Simulations with toy models are planned.  
It is conceivable that differences from standard quantum mechanics could be seen from rapidly repeated
 measurements or possibly a series of weak measurements.

\section{Conflict with Relativity?}
The concept of ``the present" is generally not considered to have a physical reality because observers in different Lorentz frames
have different present hypersurfaces in special relativity. Nevertheless the present seems like quite a special time to us which
can be seemingly be distinguished from past and future. Philosophers have struggled with the notion that we exist at the present 
and are compelled to move
through time. Some have postulated a second time to pace our motion through time\cite{dunne}. A ``growing block" theory of
time is discussed by some philosophers, which closely resembles a growing phase boundary\cite {block}.  Whitehead invented the term
``concrescence," a process through which a formless future becomes a concrete reality at the present, 
to form a fixed past\cite{whitehead}. These ideas are also explored in the already mentioned book by 
Muller\cite{muller}, who envisions
new space {\em and} time being created at the edge of the Hubble expansion. In the Phase Boundary Universe proposal, the present
is the physical surface of the growing crystal - the phase boundary itself. This is a simultaneous surface in the frame of the 
Hubble expansion. It is in this frame that the coordinate time of the present hypersurface tracks exactly the Langevin (Newtonian) time.

Ascribing reality to the present clearly
requires choosing such a 
preferred frame, an anathema to anyone trained in relativity.  Nevertheless it is possible to have a preferred frame and still get 
most if not all of what relativity provides. This will happen if interactions in the Lagrangian of the theory are all Lorentz
invariant.  Measuring rods and clocks built from matter that obeys these Lorentz invariant interactions will necessarily
behave as predicted by the Lorentz transformation when moving. In other words they will obey Lorentz contraction and time dilation.
An example of this is the Lorentz contraction (relative to the speed of sound) of the displacement field 
surrounding a moving screw dislocation\cite{screw}. This is a dynamical realization of the Lorentz symmetry as opposed to 
our usual kinematical
formulation of an empty Minkowski space to which particles are added.  The moving observer using these physical clocks and rods
will see special relativity as fully reciprocal, simply due to the property that the inverse Lorentz transformation is also a Lorentz
transformation. Despite being reciprocal, the {\em reasons} that moving rods shrink and clocks run slow are different in the
preferred frame and other frames.  From the preferred frame this is a physical effect of the dynamics of particle interactions.
From another frame looking back at the preferred frame it is more of an illusion based on the moving frame's 
use of slowed clocks and
shrunken rods for measurement, along with the different clock synchronization scheme that results from their use.
In the early days of relativity Lorentz and others still clung to the ether as a preferred frame, 
even though unobservable\cite{whittaker},
with moving rods and clocks shrinking and slowing due to their interactions with the ether.  This point of view could 
not be proven wrong because it is equivalent to special relativity, but the ether eventually fell to Occam's razor as 
an unnecessary element.

So having a preferred frame in a theory does not necessarily violate special relativity. This has been emphasized by Bell\cite{bell2}
and discussed by numerous authors\cite{preframe}. For a theory in which the Lorentz symmetry arises
dynamically, the preferred frame may be the most logical to calculate within, even if not 
experimentally distinguishable. It may simply be a calculation aid
similar to gauge fixing or choosing a coordinate system in general relativity.  Nevertheless one must also
keep aware of possible frame
dependence in the measurement process.  A measuring device has its own frame in which it is simultaneously sensitized, and this
needs to be taken into account if the device is moving relative to the preferred frame. Measuring devices on Earth, however,
are moving
only about 0.0012c relative to the Hubble expansion as measured by cosmic background radiation\cite{cbe}, so the frame
difference of most Earthbound experiments from the Hubble frame is actually rather slight and may not be of practical importance. 

It is unclear whether giving the preferred frame of the Hubble expansion a role in quantum evolution, as in this paper,
makes it observable or not. That is a subject of further study.
In the Phase Boundary Universe there is an additional feature that does make the preferred frame detectable at high energies. If
the solid phase is crystalline, the continuous translation symmetry is spontaneously broken to a discrete one.  
For inverse momenta close to the lattice spacing, the photon dispersion relation should become phonon-like which will only
be spherically symmetric in the Hubble frame. Lorentz symmetry is only approximate in this theory. A preferred frame of this sort
also opens the door a crack to possible faster than light (FTL) communication or interaction, just as a bullet can break the sound barrier.
The reason is that causality is arbitrated by the Langevin time which is equivalent to the coordinate time only in the preferred frame, 
so causality-based arguments against FTL are removed. 
Nevertheless, it still could be the case that there simply are no interactions that  break light speed.

\section{Application to Non-Relativistic Quantum Theory}

Relativistic quantum field theory is the most correct quantum theory we have, so it is really only necessary to
describe how it is to be modified to incorporate measurement, eqns. 1-3 above.  Nevertheless it would also be useful to have
a similarly-modified version of ordinary first-quantized non-relativistic quantum mechanics, since many systems are more easily
described and studied within that theory.
Non-relativistic quantum mechanics has been shown to be equivalent to a stochastic model in a number of different
but probably equivalent ways\cite{stochastic,otherstochastic}. The most straightforward approach\cite{stochastic}, however, 
has essentially
a single time with a stochastic trajectory $x(t)$ following the Langevin equation
\begin{equation}
\dot{x}= -\frac{\partial W(x)}{\partial x}+\eta (t)
\end{equation}
$W(x)$ is related to usual the quantum potential by a Ricati equation.
\begin{equation}
V=\frac{1}{2\sigma}\left[ \frac{\partial W}{\partial x}\right]^2 -\frac{1}{2} \frac{\partial ^2 W}{\partial x^2}
\end{equation} 
(the above is the Euclidean version). This approach which is more like the Nelson approach 
has even been modeled using a classical analog 
computer with a noise input\cite{asq}.
Although this scheme
gives correct results for quantum mechanics when averaged over noise histories, each trajectory considered has a fixed history.
The two-time approach advocated
above, however, requires a fluctuating history within each trajectory.  One needs a model where the 
entire history functional $x(t)$ 
fluctuates in the 
second time $t'$.  This may be achieved 
through the path integral approach, which is the same as the field theory approach given above with $\phi(x,t)$ replaced with $x(t)$.
The associated Langevin equation is  
\begin{equation}
\frac{\partial x(t,t')}{\partial t'}= i\frac{\partial S(x)}{\partial x} \bigg|_{x=x(t,t')} +\eta (t,t')-\epsilon x(t,t')
\end{equation}
Since the particle's worldline does not exist beyond the present hypersurface, new variables $x(t=t')$ must be introduced
as $t'$ evolves. In other words the path ends at the present, $t=t'$. This feature is quite different from the standard treatment
in which the Langevin time has no relationship to the coordinate time.
Measurement is either modeled explicitly as with the field-theory case, or paths for the propagator
$<x(t_1 )x(t_2 )>$
are averaged over a measurement time $\Delta$ as in Eqn. \ref{this}.  For a fixed initial condition, $x(t_1 )$ would not vary
with $t'$, but all other $x(t)$ would. Further averaging over different noise histories would be done with the 
squared propagator or whatever probability was being calculated, because a measurement process is being
modeled and each measurement terminates an experiment.  Multiple experiments generate a probability distribution as in ordinary
statistics. Simulations with simple quantum systems are planned to explore the differences between this proposal and 
standard quantum mechanics. 

\section{Conclusion} 
 
The hypothesis presented in this paper is that the Langevin time of stochastic quantum quantization is a physical time over which
quantum fluctuations take place and also by which particles move through {\em both} space and coordinate time. This allows 
quantum measurements to be modeled as a process of spontaneous symmetry breaking, even in a finite system. Quantum
fluctuations are averaged over time during the measurement process which effectively implements quantum superposition. One 
advantage of this approach is that measurement is built into the theory - no separate measurement process need be added on.
This makes resolution of quantum paradoxes particularly straightforward. Because of its close relationship to standard stochastic
quantization, this theory is expected to closely mimic standard quantum theory, although some differences could exist. These
will be explored in future work with simple test systems.
Since the measurement process is built in, the potential to construct
computational engines with the power of a quantum computer seems possible using such a
noisy classical system with one added dimension (the second time).
To have multiple bits at least a 2-d array of noise generators is needed.
However, this could also just end up reproducing a known algorithm such as simulated annealing applied to a system
resembling to a spin glass. The analog version of simulated annealing would be annealing itself. In this case ``thermal computing"
would rival quantum computing.  A better understanding of the
relationship of quantum systems to noisy classical systems will likely generate insights into both the capabilities and
limitations of quantum computers.

\newpage


\begin{thebibliography}{99}
\bibitem{parisi-wu}Parisi, G., Wu, Y.S.: Perturbation theory without gauge fixing. Sci. Sin. 24, 483-496 (1981)
\bibitem{sqmin}H\"{u}ffel, H., Rumpf, H.: Stochastic quantization in Minkowski space. Phys. Lett. 148B, 104-110 (1984); 
Nakazato, H., Yamanaka, Y.: Minkowski stochastic quantization. 
Phys. Rev. D 34, 492-496 (1986); Nakazato H.: Thermal equilibrium in Minkowski stochastic quantization.
Prog. Theor. Phys. 77, 20-25 (1987)
\bibitem{muller}Muller, R.A.: Now - The Physics of Time. W.W. Norton, New York (2016); Muller, R.A., Maguire, S.: Now,
and the flow of time. ArXiv:1606.07975v1 (2016).
\bibitem{pbu}Grady, M.: Quantum mechanics, quantum gravity, and approximate Lorentz invariance from a 
classical phase-boundary universe. In: Moore D.C. (ed.) Trends in Quantum Gravity Research, pp. 109-137 Nova Science Pub., 
New York (2006); Grady, M.: Universe as a phase boundary in a four-dimensional Euclidean space. ArXiv:gr-qc/9805076 (1998); 
Chown M.: Cosmic crystal. New Scientist 161, 42 (1999)
\bibitem{kleinert}Kleinert, H., Zaanen, J.: Nematic world crystal model of gravity explaining absence
of torsion in spacetime. Phys. Lett. A 324, 361-365 (2004); Kleinert,H.: Gravity as a theory of defects in
a crystal with only second-gradient elasticity. Ann. der Phys.  44, 117-119 (1987)
\bibitem{volovik}Volovik, G.E.: 3He and universe parallelism. In: Bunkov, Y.M., Godfrin, H.(eds.) 
Topological Defects and the Non-Equilibrium Dynamics of Symmetry Breaking Phase Transitions, pp.353-387. 
Kluwer Academic Publishers, Dordrecht Boston (2000); Volovik, G.E.:
The Universe in a Helium Droplet. Oxford University Press, Oxford (2009) 
\bibitem{ghirardi}e.g. Pearle, P.: Reduction of the state vector by a nonlinear Schr\"{o}dinger equation.
Phys. Rev. D 13, 857-868 (1976); Ghirardi, G.C., Rimini, A., Weber, T.: Unified dynamics for microscopic and macroscopic systems.
Phys Rev. D 34, 470 (1986); Ghirardi, G.C., Pearle, P., Rimini, A.: Markov processes in Hilbert space and 
continuous spontaneous localization of systems of identical particles. Phys. Rev. A 42, 78-89 (1990)
\bibitem{microcanonical}Gross, D.H.E.: Microcanonical Thermodynamics. World Scientific, Singapore (2001)
\bibitem{qmasssb}Ne'eman, Y.: Problems in quantum foundations in light of gauge theories. 
Found. Phys. 16, 361 (1986);
Zim\'{a}nyi, G.T., Vlad\'{a}r, K.: Possible role of symmetry breaking in quantum measurement theory.
Phys. Rev. A 34, 3496 (1986), Symmetry breaking and measurement theory. Found. Phys. Lett. 1, 175, (1988); 
Grady, M.: Spontaneous symmetry breaking as the mechanism of quantum measurement. ArXiv:hep-th/9409049 (1994) 
\bibitem{decoherence}e.g. see Griffiths, R.B.: Consistent Quantum Theory. Cambridge University Press, Cambridge (2002)
\bibitem{adler}Adler S.L.: Why decoherence has not solved the measurement problem: a response to P.W. Anderson.
 Stud. Hist. and Phil. of Mod. Phys. 34, 135-142 (2003)
\bibitem{epr}Einstein, A., Podolsky, B., Rosen, N.: Can quantum-mechanical description of reality be considered complete? 
Phys. Rev. 47, 777-780 (1935)
\bibitem{bell}Bell, J.: On the Einstein-Podolsky-Rosen paradox. Physics, 1, 195-200 (1964)
\bibitem{aspect}Freedman, S.J., Clauser, J.F.: Experimantal test of local hidden variable theories. Phys. Rev. Lett. 28, 938-941 (1972); 
Aspect, A., Grangier, P. and Roger, G.: Experimental realization of 
Einstein-Podolsky-Rosen-Bohm gedankenexperiment: a new violation of Bell's inequalities. Phys. Rev. Lett., 49, 91-94 (1982)
\bibitem{noFTL}e.g. Eberhard, P.H., Ross, R.R.: Quantum field theory cannot provide faster-than-light communication. Found.
Phys. Lett. 2, 127-149 (1989); Peres, A., Terno, D.R.: Quantum information and relativity theory. Rev. Mod. Phys. 76, 93-123 (2004)
\bibitem{neeman}Ne'eman, Y.: Classical geometric resolution of the Einstein-Podolsky-Rosen paradox. 
Proc. Natl. Acad. Sci. USA  80, 7051-7053 (1983) 
\bibitem{ev}Elitzur, A.C., Vaidman L.: Quantum mechanical interaction-free measurements. Found. Phys. 23, 987-997 (1993)
\bibitem{stochastic}Schneider, T., Zannetti, M., Badii, R.: Stochastic simulation of quantum systems and critical dynamics. 
Phys. Rev B  31, 2941-2951 (1985)
\bibitem{otherstochastic}Klauder, J.R.: Coherent state Langevin equations for canonical quantum systems with 
applications to the quantized Hall effect. Phys. Rev. A 29, 2036-2047 (1984)
\bibitem{kleinert2}Haba, Z., Kleinert, H.: Schr\"{o}dinger wave functions from classical trajectories.
Phys. Lett. A  294, 139-142 (2002)
\bibitem{mueller}Biro, T.S., Matinyan, S.G., M\"{u}ller, B.: Chaotic quantization of classical gauge fields. Found. Phys. Lett. 14,
471-485 (2001)
\bibitem{nelson}Nelson, E.: Derivation of the Schr\"{o}dinger equation from Newtonian mechanics. Phys. Rev. 150, 1079-1085 (1966)
\bibitem{asq}Stocks, N.G., Lambert, C.J., McClintock, P.V.E.: Analogue simulation of quantum mechanical systems.
J. Stat. Phys. 54, 1397-1410 (1989)
\bibitem{dunne}Dunne, J.W.: An Experiment with Time. Faber and Faber, London (1938)
\bibitem{block}Broad, C. D.: Scientific Thought. Harcourt Brace and Co., New York (1923)
\bibitem{whitehead}Whitehead, A.N.: Process and Reality. Macmillan, New York (1929)
\bibitem{screw}Frank, F.C.: On the equations of motion of crystal dislocations. Proc. Phys. Soc. A62, 131-134 (1948); 
Weertman, J., Weertman, J.R.: Elementary Dislocation Theory. Oxford University Press, Oxford (1992)
\bibitem{whittaker}Whittaker, E.: A History of the Theories of Aether and Electricity, Vol.II: The Modern Theories. 
Thomas Nelson and Sons, London p.36, (1953)
\bibitem{bell2}Bell, J.S.: Speakable and Unspeakable in Quantum Mechanics. Cambridge Univ. Press, Cambridge (1988)
\bibitem{preframe}e.g. Guerra, V., de Abreu, R.: On the consistency between the assumption of a special system of 
reference and special relativity. Found. Phys. 36, 1826-1845 (2006) and references therein
\bibitem{cbe}Fixsen, D.J. et. al.: Cosmic microwave background dipole spectrum measured by the 
COBE FIRAS instrument. Ap. J. 420, 445-449 (1994) 


\end{thebibliography}
\end{document}